# HyperSmooth : calcul et visualisation de cartes de potentiel interactives.

**Christine Plumejeaud**[(1)], **Jean-Marc Vincent**[(1)], **Claude Grasland**[(2)], **Jérôme Gensel**[(1)], **Hélène Mathian**[(2)], **Serge Guelton**[(1)], **Joël Boulier**[(2)]

*(1) Laboratoire d'Informatique de Grenoble, BP 72  38402   Saint-Martin d'Hères*
   *jérôme.gensel@imag.fr*

*(2) UMR Laboratoire Géographie-Cités,  13 rue du four,  75006 Paris*
   *claude.grasland@parisgeo.cnrs.fr*

RÉSUMÉ. *Le projet de recherche Hypercarte propose un nouvel outil cartographique pour l'analyse spatiale de phénomènes sociaux économiques mettant en œuvre une méthode de calcul de potentiel. L'objectif étant de pouvoir représenter de façon continue et en changeant d'échelle d'analyse une information statistique échantillonnée sur des maillages administratifs. Le défi technologique est de fournir un outil accessible via le Web, interactif et rapide, ceci malgré le coût élevé du calcul, et qui assure la confidentialité des données. Nous présentons notre solution, ses résultats, et les améliorations que nous pourrions lui apporter.*

ABSTRACT. *The HyperCarte research group wishes to offer a new cartographic tool for spatial analysis of social data, using the potential smoothing method. The purpose of this method is to view the spreading of phenomena's in a continuous way, at a macroscopic scale, basing on data sampled on administrative areas. We aim to offer an interactive tool, accessible via the Web, but guarantying the confidentiality of data. The major difficulty is induced by the high complexity of the calculus, working on a great amount of data. We present our solution to such a technical challenge, and our perspectives of enhancements.*

MOTS-CLÉS : *cartographie – analyse spatiale – informatique –  lissage multiscalaire – interface utilisateur –  parallélisation – web services – SOAP – quadtree.*

**1. Introduction**

Les avancées dans le domaine du Web ont ouvert de nouvelles perspectives dans le domaine de la cartographie interactive et dynamique. Dans ce contexte, le groupe de recherche multidisciplinaire HyperCarte[1] s'est donné pour objectif de concevoir et implémenter une collection cohérente de plates-formes interactives d'analyse spatiale et de représentations cartographiques de phénomènes sociaux, économiques, environnementaux, etc. Les applications visées se situent principalement dans le domaine socio-économique ou environnemental et l'aide à la décision en matière de prospective territoriale pour un public large : chercheurs en géographie, en sciences sociales et humaines, mais aussi décideurs politiques et grand public. L'enjeu est de proposer une cartographie s'adressant à des utilisateurs aux compétences diverses, néophytes ou spécialistes. L'évolution des technologies associées au Web offre d'énormes possibilités à ce type d'approche, en particulier grâce à la souplesse qu'apporte une interactivité de haut niveau.

L'objet de cet article concerne plus précisément la réalisation d'un environnement basé sur l'infrastructure du Web capable de générer dynamiquement des cartes continues, c'est-à-dire affranchies d'éventuels maillages d'observation (administratifs, grille de collecte, etc.). L'objectif est de visualiser la distribution spatiale des phénomènes analysés à un niveau macroscopique, c'est-à-dire largement supérieur aux maillages d'observation. La méthode que nous proposons, appelée méthode de transformation par potentiel, conserve la masse totale des données et en donne une représentation non biaisée. Cette méthode, lorsque la portée du potentiel est petite, peut être appliquée à du lissage de données.

Une des difficultés de diffusion de cette méthode d'analyse spatiale et de représentation cartographique vient d'une part du coût élevé du calcul[2] qui empêche de répondre à l'exigence d'interactivité de l'utilisateur, et d'autre part de la contrainte sur la sécurité et la confidentialité des données analysées.

Cet article présente la solution conçue pour répondre à ces besoins. Elle s'est développée suivant trois axes : (i) distribution des calculs sur un serveur multi-processeur détenant les données, (ii) parallélisation des tâches de calcul, (iii) visualisation des cartes sur un client web interactif avec connexion sécurisée.

Cette présentation s'organise autour du plan suivant : dans un premier temps, les fondements théoriques et les applications possibles de la méthode du potentiel sont rappelés, et cette méthode est située en regard de ce qui existe. Suivent ensuite la justification de l'architecture retenue et la description de la stratégie d'optimisation des calculs. Enfin, les résultats de cette réalisation sont présentés ainsi que les perspectives d'amélioration qu'elle offre.

---

[1] http://www-lsr.imag.fr/HyperCarte/
[2] Il faut plusieurs heures pour calculer une carte de résolution moyenne visualisant la densité de population en France, à partir du recensement sur les 36000 communes.

**2. La méthode de transformation par potentiel**

*2.1 Principe*

Les représentations cartographiques continues de phénomènes spatiaux discrets sont nécessaires lorsque l'on veut s'abstraire d'un maillage spatial, soit parce que le maillage est hétérogène ou qu'il n'est pas signifiant pour le phénomène étudié, afin de ne garder que l'organisation spatiale du phénomène, sans référence au découpage sous-jacent du territoire. Du point de vue méthodologique, elle s'apparente aux méthodes de traitement du signal par déconvolution du signal échantillonné.

La méthode consiste à considérer un espace géographique donné, sur lequel est plaqué un maillage constitué d'unités territoriales et de leur stock associé, puis à calculer en tout point de l'espace discrétisé la valeur du potentiel de chaque stock. La discrétisation est une division de l'espace en parcelles régulières, via la projection d'une grille par exemple.

Soit $A$ l'ensemble des unités territoriales, $a$ un élément de cet ensemble, $Sa$ la valeur du stock sur cette unité. Sachant que les effets des stocks s'additionnent et sont liés à la distance $\delta$ entre $a$ et le point $M$, nous définissons le potentiel $\Phi(M)$ en tout point $M$ de l'espace géométrique par :

$$\Phi(M) = \sum_{a \in A} S_a f(\delta(a,M)) . \qquad [1]$$

Par exemple, si $A$ est l'ensemble des communes européennes (ou équivalents dans le NUTS 5), $a$ la commune de Nuoro en Sardaigne, alors $Sa$ serait le nombre de personnes centenaires vivant à Nuoro.

On spécifie la distance $\delta(a,M)$ en mesurant l'éloignement $d$ entre $M$ et $g_a$, un point représentatif de $a$ (qui peut être son centre de géométrie, son centre administratif ou industriel, etc.). La contribution au potentiel de chaque élément $e$ est pondérée par une fonction $f$ de la distance $d$, car l'effet d'un stock diminue usuellement avec la distance, et il est maximal à une distance nulle, et nul à une distance infinie. Pour que le potentiel ait du sens, en particulier lorsqu'il dépendra d'un paramètre, on normalise celui-ci par l'équation [2], en tout point O de l'espace:

$$\int_{R^2} f(d(O,M)).dM = 1 \text{, soit encore } \int_A \Phi(M).dM = \sum_{a \in A} S_a . \qquad [2]$$

La somme totale des stocks est égale à l'intégrale du potentiel, on obtient ainsi une redistribution de la masse sur l'espace considéré.

Dans une métaphore du modèle de gravité, $\Phi(M)$ s'interprète comme l'attraction exercée par l'environnement sur un mobile placé en M, dont le vecteur de déplacement serait alors –grad $\Phi$. Une interprétation duale serait aussi que $\Phi(g_a)$ mesure l'influence d'une masse placée en $g_a$ sur l'ensemble des points M de son voisinage.

De l'équation [1], il découle que la complexité du calcul dépend à la fois de la taille de l'espace administratif (le nombre *n* d'éléments *a*), et de la résolution de l'image à produire (le nombre *m* de points *M* que l'on estime).

Le calcul dépend principalement de la fonction *f*, **fonction d'interaction spatiale.** Elle intègre les hypothèses concernant les lois de diffusion dans l'espace associées au phénomène étudié. Trois modèles de fonctions paramétrées sont proposés : un modèle à support limité (disque et disque amorti), un modèle exponentiel (Gaussienne) pour des interactions proches, la décroissance de *f* se faisant selon une exponentielle négative, et enfin un modèle à interaction de longue portée (Pareto), la décroissance suivant une puissance inverse. Par exemple, cette méthode permet de modéliser et d'étudier la propagation des épidémies touchant l'homme : leur diffusion pourra se faire soit sur de longues distances, soit sur de courtes distances, suivant le rayon d'action de l'élément contaminant (Grasland *et al.*, 2005-a). L'utilisateur peut tester différents modèles en choisissant la fonction d'interaction à appliquer.

L'analyse du phénomène dépend aussi de la **portée** *p* de la fonction d'interaction. La portée est définie comme la distance moyenne d'action d'une masse sur son voisinage. Elle est reliée à la forme de la fonction d'interaction par l'équation suivante en tout point *O* de l'espace:

$$p = \int_{R^2} d(O,M) f(d(O,M)).dM = \int_0^{+\infty} f(r) 2\pi r^2 .dr. \qquad [3]$$

La portée peut être interprétée comme l'échelle spatiale de représentation choisie. Le tandem (fonction, portée) traduit concrètement les hypothèses économiques et sociologiques associées aux interactions entre les acteurs sur le territoire. A portées identiques, le calcul du ratio de deux potentiels de stocks différents s'interprète comme une densité. Par exemple, le potentiel $\Phi_P$ de population peut se rapporter au potentiel $\Phi_S$ de superficie, ce qui nous donne une densité de population sur une portée donnée et uniforme.

De plus, les densités calculées peuvent se comparer sur des portées différentes. Ainsi, une étude sur la structure des inégalités régionales en Europe (Grasland *et al*., 2007) a mis en évidence la structure polycentrique du peuplement européen en comparant les densités de population à différentes portées de lissage. Une densité locale de 200 hab/km2 pour une portée $P_1$ correspondra à un « pic » de population si la densité est plus faible à la portée $P_2$ ($P_2 > P_1$) ou au contraire à un « creux » si la densité à la portée du voisinage $P_2$ est plus forte. Ce type de comparaison permet d'étudier la situation d'un territoire dans un voisinage plus ou moins étendu, et permet d'aller jusqu'à une modélisation des flux associés aux différentiels locaux (Grasland, 2003). Il est donc important de pouvoir paramétrer la portée et qu'elle s'exprime en terme de distance.

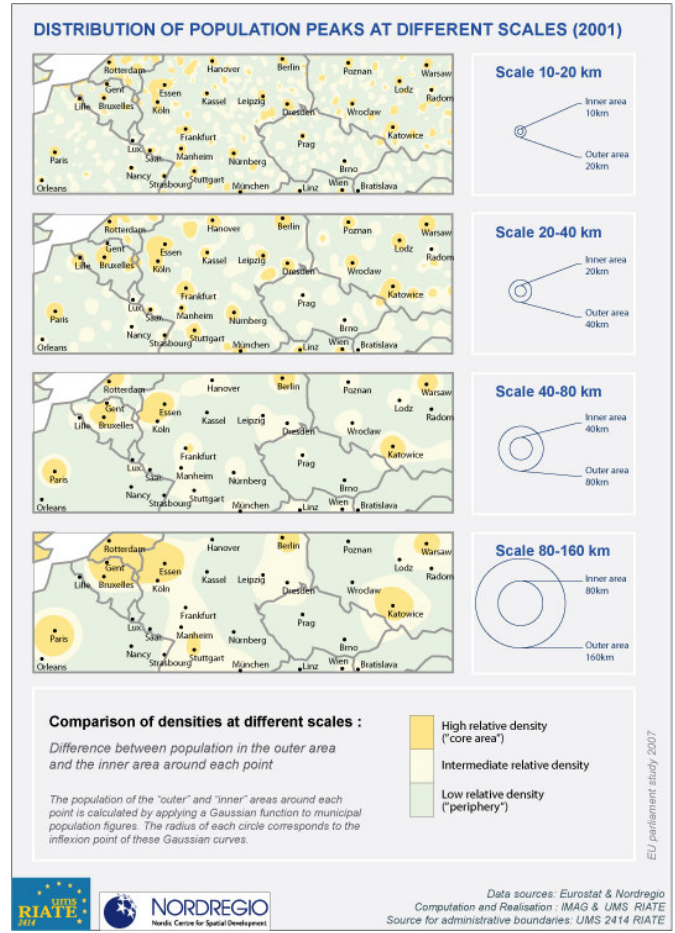

**Figure 1.** *Analyse multiscalaire des pics de densités de population en Europe. Source (Grasland et al, 2007).*

D'autre part, il faut choisir le type de **distance** utilisée. Elle dépend de la taille de l'espace couvert par la carte. Par exemple, à l'échelle d'un continent, on ne peut pas utiliser la distance euclidienne sans introduire des déformations. La distance orthodromique convient alors mieux car elle tient compte de la sphéricité de la Terre. En outre, le type de phénomène analysé peut nécessiter parfois d'utiliser des distances tenant compte de l'anisotropie de l'espace (distance temps voiture par exemple). Mais notre proposition actuelle se limite à la distance orthodromique.

*2.2 Quels usages ?*

L'intérêt de la méthode a d'ores et déjà été documenté par les géographes (Grasland, 2003), et des expériences dans différents domaines ont prouvé l'apport de telles représentations (Lacaze, 2000), (Dumas, 2001), (Guérois, 2003)**.** Un grand nombre de phénomènes socio-économiques, et surtout environnementaux, obéissent à des logiques de diffusion et de distribution dans l'espace relevant de fonctions continues, sans extinction brutale le long des frontières.

Par exemple, une cartographie des implantations de type Seveso dans le cadre d'unités administratives de petite taille n'a guère de sens puisque les pollutions éventuelles se propagent de façon continue dans l'espace et ne respectent évidemment pas les limites politiques ou administratives. La cartographie correcte consiste à déterminer la portée spatiale des émanations dangereuses (par exemple une fonction exponentielle négative décroissant de 0 à 100 km), puis à calculer en tout point de l'espace le potentiel d'exposition à une pollution de type Seveso.

La même remarque vaut lorsque l'on étudie le taux de chômage dans un marché du travail non segmenté spatialement (libre circulation des travailleurs) : la meilleure représentation consiste à partir de la distance temps maximale pour se rendre au travail (par exemple, 2 heures de temps routier), puis, à calculer, en tout point de l'espace, le taux de chômage dans un voisinage de 2 heures. L'originalité de ce dernier exemple réside dans le fait qu'un accroissement des mobilités (amélioration des conditions de transport) modifie la cartographie du chômage, ce qui n'est évidemment pas pris en compte dans une cartographie classique par département ou bassin d'emploi.

Outre la restitution de la continuité de phénomènes sociaux, cette méthode a prouvé ses qualités aussi pour :

- Etudier des évènements rares : en ajustant la portée finement, les différenciations spatiales sont conservées tout en évitant d'interpréter des maxima non significatifs. Un bon exemple est fourni par la contribution à l'étude de la « zone bleue » en Sardaigne où la visualisation des potentiels sur une portée courte met en évidence une zone de présence étonnamment forte de centenaires (Poulain et al, 2004).

**-** Filtrer des distributions complexes : une étude de la répartition de l'occupation des sols sur le territoire français en est une illustration. Réalisée en 1998-1999 par l'IFEN, elle dégage ainsi des continuités par types d'occupation des terres et visualise les compétitions entre des usages divers. La méthode a été ensuite transposée à l'échelon européen dans le cadre du projet CORILIS. (Lacaze et al, 2000)

- Représenter des données confidentielles ou sensibles : employée pour étudier les inégalités scolaires dans l'agglomération parisienne à l'aide de données originales sur les collèges, la méthode montre comme autre atout qu'elle préserve la confidentialité des données (en localisant sans stigmatiser un collège particulier), (François J-C, 1996)

*2.3 Situation par rapport à d'autres méthodes*

La méthode des potentiels s'inscrit dans la lignée des méthodes proposées pour solutionner les problèmes d'instabilité des résultats, notamment les corrélations, en fonction de la résolution spatiale choisie. Ces problèmes identifiés comme le *Modifiable Areal Unit Problem* (MAUP) ou plus généralement le *Change Of Support Problem* (COSP), ont donné lieu à divers développements méthodologiques et discussions sur leurs apports et contraintes respectives [Gotway et al., 2002]. L'objet de cet article n'est pas de revenir spécifiquement sur ces discussions, ce qui a été fait par ailleurs (Grasland, et al, 2006). Nous proposons cependant de la situer par rapport aux méthodes les plus répandues et de pointer ici quelques unes de ses caractéristiques.

Elle est conceptuellement très différente des méthodes d'interpolation telles que triangulation, Krigeage simple, moyenne locale, ou méthode de Shepard, qui, si elles sont adaptées à l'estimation de variables continues dans l'espace pour lesquelles il existe des points de mesures (comme la température, ou l'altitude), ne le sont pas pour des variables résultant d'un comptage sur une zone délimitée à l'intérieur de laquelle la répartition effective de la population est inconnue (Grasland, et al, 2006). Ainsi elles ne font pas disparaître le maillage sous-jacent, et au contraire, elles pourraient faire croire que les disparités observées sont le fait du phénomène étudié, alors qu'en réalité elles sont la résultante de l'hétérogénéité du maillage.

Elle se distingue aussi des méthodes géostatistiques qui comme le krigeage par bloc analysent la répartition spatiale des données par des mesures de variance, et utilisent ces résultats pour inférer une surface continue. La méthode du potentiel propose un modèle de diffusion a priori défini par le couple (fonction, portée).

La méthode la plus proche est la méthode pycnophylactique (Tobler, 1979), qui elle aussi conserve la masse des données. Elle réalloue les stocks mesurés à l'intérieur de chaque maille de façon à : (i) garantir la continuité avec une maille voisine, (ii) conserver sur chaque maille la masse mesurée. En raison du prédicat (ii), la visualisation dépend alors du maillage utilisé pour l'analyse : les résultats sont différents selon le niveau d'agrégation des données, alors que ce n'est pas le cas avec le potentiel.

Cependant, la méthode de transformation par potentiel ne résout pas certains problèmes rencontrés aussi dans d'autres méthodes. Les données sont mesurées sur un espace limité, sur les bordures du duquel la méthode ne peut pas fonctionner de façon isotrope. L'étendue des marges présentant un défaut d'évaluation est directement proportionnelle à la portée. De plus, en dessous d'une certaine portée, la méthode devient imprécise ; la portée minimale peut être calculée par l'analogue du théorème de Nyquist : elle vaut deux fois la taille maximale des mailles (Nyquist, 1928). Enfin, un maillage trop hétérogène conduit à faire un compromis entre les portées minimum associées à chaque classe de taille d'unité.

**3. Proposition technique**

Le volume de données à traiter et les ressources exigées pour les calculs sont, de fait, importants. Aussi, nous déportons la partie graphique avec la visualisation du coté d'un client Web Java, tandis que les calculs et le traitement lourd de données sont effectués sur un serveur accessible à distance, via le protocole SOAP (*Simple Object Access Protocol)* avec sécurisation des échanges via SSL (*Secure Socket Layer*). Le traitement des données coté serveur est optimisé de façon à produire des résultats intermédiaires dans des temps n'excédant pas quelques secondes.

*3.1 Communication entre le serveur et le client*

Le format des données renvoyées par le serveur est contraint par les fonctionnalités attendues sur le client. Outre l'interactivité sur le choix des palettes, du nombre de classes et du type de progression dans la distribution des couleurs, on souhaite que le client puisse générer à la demande un rapport numérique (sous forme de fichier texte ou bien HTML) contenant les coordonnées géographiques de chaque point *M*, avec la valeur de son potentiel. Le client a donc tout avantage à récupérer et sauver la grille matricielle des valeurs de potentiel calculées plutôt qu'une image au format PNG, JPEG, ou GIF. Il peut ainsi redessiner une image rapidement en cas de changement de préférences graphiques.

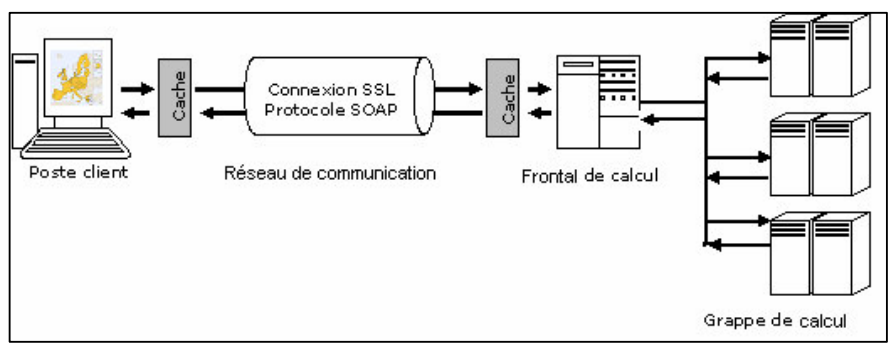

**Figure 2.** *Vue générale de l'infrastructure distribuée.*

La grille de potentiels transmise en valeurs flottantes non tronquées que calcule le serveur dépend des paramètres de calcul (résolution, cadrage, fonction d'interaction et portée) transmis par le client. La nature et le domaine de validité de ces paramètres sont établis par un contrat commun partagé par le serveur et le client.

*3.2. Stratégie d'optimisation*

Remarquons, en premier lieu, que les stratégies classiques de gestion de cache ne sont pas adaptées à notre cas car chaque requête doit générer un résultat global qui ne peut être pré-calculé puisqu'il dépend des paramètres de l'analyse. Cependant un examen approfondi des tâches de calcul sur le serveur montre qu'il existe des redondances que nous pourrions exploiter pour optimiser certaines parties du calcul. (Grasland, 2005). En effet, deux problèmes se posent lorsque nous calculons le potentiel *Φ(M)* (confère l'équation [1] ): d'une part, la somme se fait sur un nombre important d'éléments, et, d'autre part, le calcul des distances orthodromiques[3] *d(M, ge)* est coûteux puisqu'il fait intervenir des calculs d'angles en arccosinus, cosinus et sinus.

Nous réduisons la somme en pratiquant une politique d'élagage (*cut-off*) algébrique. Contrairement au *cut-off* géométrique des géographes qui limite le calcul sur un certain rayon de portée, nous tenons compte de points éloignés dont le poids *Se* (valeur statistique) influence le résultat du calcul. Notre algorithme de *cut-off* utilise une méthode numérique par recherche dans un arbre (*quadtree*).

A chaque feuille de l'arbre de profondeur *n* est associé un point *M* valorisé avec ses coordonnées et son stock. Chaque point est ensuite sommé par groupe de 4 (des voisins sur la grille). Le calcul du potentiel *Φ(M)* s'effectue récursivement en cumulant le produit des stocks associés aux feuilles, par leur distance au point *M*.

La visite de chaque branche de niveau *n-1* dépend de la réussite du test suivant, qui vérifie si le poids des enfants du nœud $_{n-1}$ est négligeable ou non :

$$( \sum_{noeud_{n-1}} S_e) * d_{\min} \leq \varepsilon * \Phi(M)_{cumulé} .\qquad[4]$$

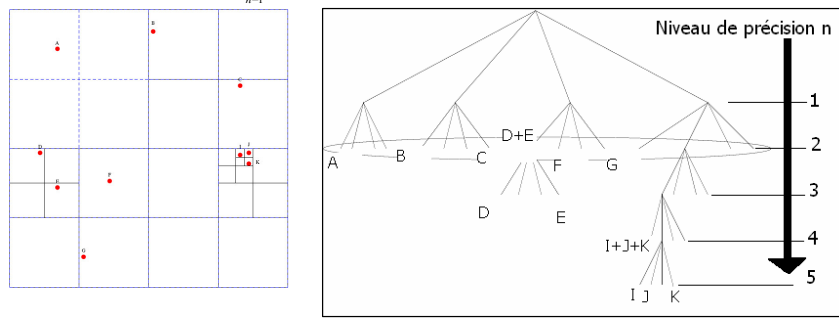

**Figure 3.** *Exemple de construction de l'arbre à partir d'un espace d'étude discrétisé.*

---

[3] Rappel : la distance orthodromique entre 2 points A et B de coordonnées respectives A(a1, a2) et B(b1, b2) avec r rayon terrestre et les latitudes, longitudes en radians vaut :
d(A,B) = arccos(sin(latA) sin(latB) + cos(latA) cos(latB) cos(longB-longA)) * r.

Le point délicat est d'ajuster la valeur de l'epsilon de manière à ne pas négliger trop de points. Par défaut, la valeur est fixée à un 1/1000 de la somme totale des stocks.

Le second facteur de lenteur de calcul (l'évaluation d'expressions contenant des termes en arccosinus, cosinus et sinus d'angles) peut être contourné en tabulant de façon fine ces fonctions. C'est-à-dire que les valeurs des fonctions sur des angles correspondant à une division régulière et fine d'un intervalle *I* donné sont pré-calculées, et pour tout angle la valeur de la fonction est approximée par la borne inférieure de la division à laquelle il appartient. Le grain de cette tabulation est fixé à l'avance par un paramètre du programme.

### 3.3. *Visualisation interactive*

L'utilisateur peut agir sur les paramètres suivants de l'analyse :

- la fonction d'interaction : une liste déroulante présente la liste des fonctions implantées sur le serveur, au format textuel que ce dernier retourne au client.

- la portée moyenne en kilomètres de l'analyse, sélectionnée via une barre de glissement, ou par saisie numérique. Ce choix n'est pas borné.

- la résolution de visualisation définie en nombre de points en largeur et hauteur désirés pour la grille, et qui se rapporte au pas de discrétisation de l'espace étudié.

- le cadrage (= le territoire ciblé) pour l'analyse, correspondant actuellement à l'espace visualisé. Un zoom ou un déplacement modifie donc le cadrage.

- les stocks proposés dans deux listes déroulantes séparées, une pour le numérateur, l'autre pour le dénominateur, afin de pouvoir ensuite calculer un ratio de potentiel. Le client interroge le serveur lors de l'authentification de l'utilisateur pour connaître la liste des stocks disponibles.

Une collection d'onglets est créée, chacun contenant une carte spécifique : le premier présente simplement l'aire d'étude et le maillage administratif. Les trois suivants présentent pour une portée $V_1$ donnée le potentiel du numérateur, du dénominateur et leur ratio $Z_1$. Les trois autres suivants de même mais sur une portée différente $V_2$. Enfin le dernier onglet compare les densités entre les deux voisinages en visualisant leur différence : $Z_2 - Z_1$.

L'utilisateur dispose en plus de fonctions de zoom, déplacement, et de navigation dans l'atlas de cartes ainsi produit. La visualisation d'une carte de potentiel se fait en superposition avec un fond vectoriel présentant les limites administratives de l'espace analysé. Ces cartes présentent une gradation de couleurs, suivant l'intensité du phénomène, paramétrable indépendamment pour chaque onglet, via une proposition à gauche de la zone de visualisation pour le choix de palette, le type de progression, et le nombre de classes dans la distribution.

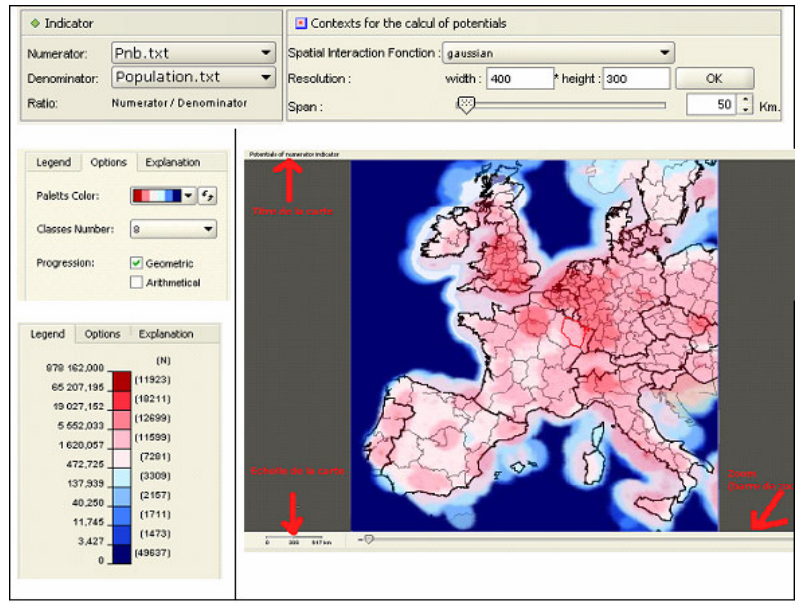

**Figure 4.** *Aperçu de l'interface, avec un ratio du PNB par habitant sur un calcul gaussien de portée 50 km.*

## 4. Bilan de la réalisation

### 4.1. Performances

A l'avenir, ce programme sera exécuté par une machine multi-processeurs de type SMP, avec mémoire partagée, et dès lors, les temps de calculs cités seraient divisés par 4, 8, ou plus, selon le nombre de processeurs mobilisés. Le serveur peut s'utiliser de façon autonome par rapport au client, et les performances suivantes (plutôt bonnes) sont des temps de calcul de grilles mesurées localement sur un serveur bi-processeurs sous Linux (Pentium 4 à 2,6 Ghz, avec 1 Go de mémoire).

| Nombre d'échantillons | Résolution | Portée (km) | Temps de calcul (s) |
|---|---|---|---|
| 116203 | 400 x 300 | 100 | 35 |
| 288 | 400 x 300 | 100 | 9 |
| 116203 | 800 x 600 | 100 | 165 |
| 116203 | 400 x 300 | 50 | 15 |

**Tableau 1.** *Temps de calcul des matrices sur le serveur*

Les temps de latence réseau sont bons, malgré le fait que nous n'avons pas encore mis en œuvre la compression des données échangées via gzip. En fait, on mesure un temps correspondant à l'emballage de la réponse, l'encryptage et le déballage de 4 s localement avec une résolution de 300 * 400. Coté client, la reconstruction de l'image à partir d'un tableau de flottants prend très peu de temps (128 ms). Cette matrice est re-calculée dès lors que l'utilisateur change un des paramètres d'analyse, ou bien agrandit, réduit ou se déplace dans la zone de visualisation, et sinon, elle est cachée avec les paramètres de la requête correspondante afin de pouvoir l'exploiter ultérieurement. La résolution demandée étant le plus souvent inférieure à celle de l'image vectorielle (1027 * 688), une interpolation bi-quadratique est appliquée sur l'image produite afin de la caler sur le fond vectoriel.

*4.2. Perspectives d'amélioration*

Plusieurs pistes s'offrent pour améliorer les performances du serveur de calcul. Par exemple le calcul de la distance orthodromique peut bénéficier d'un pré-calcul supplémentaire. Sans entrer dans les détails, il faut encore exploiter les travaux de recherche concernant la quête du compromis idéal entre consommation d'espace mémoire et gain de temps obtenu par tabulation. De même, la troncature des flottants transmis dans la grille matricielle réduirait le volume des données échangées, et accorderait donc un gain de temps sur la transmission des données. Un algorithme qui automatise la détermination du niveau de troncature est en cours d'élaboration.

D'autres pistes, moins classiques, sont liées à l'usage de notre algorithme de *cut-off*. Il s'agit de préparer une stratégie de sous-échantillonage des données : par exemple, adapter la valeur du seuil d'élagage dynamiquement au lieu de fixer arbitrairement sa valeur ; ou bien limiter la visite de l'arbre à un certain niveau, sans aller jusqu'aux feuilles, lorsque que le niveau d'information requis est très grossier.

Enfin, la sélection de la distance pourrait se faire de manière adaptative, selon l'échelle d'étude retenue. Ainsi, employer la distance euclidienne sur la région Rhône-Alpes n'introduirait pas de déformations notables.

De façon identique, quelques améliorations pour le client sur le plan de l'ergonomie sont prévues. Comme, par exemple, décorréler la zone de visualisation et la zone de traitement, en utilisant un cadre de sélection de la surface à traiter. Nous souhaitons donner une interprétation plus directe de la résolution : indiquer un pas en nombre de kilomètres, plutôt qu'un nombre de points sur une grille. Enfin, le calcul et la représentation de courbes de niveau à partir de l'image matricielle amélioreraient sensiblement le fondu avec le fond vectoriel représentant le maillage administratif.

Le client offre aussi un champ de réflexions en ce qui concerne l'usage de caches autorisant un raffinement progressif des images, grâce à l'émission de requêtes successives. En effet, tant que l'utilisateur ne modifie pas ses souhaits, il serait judicieux de construire une image dont la résolution augmente dans le temps, et d'anticiper un zoom sur la zone lissée.

## 5. Conclusion

Cet article présente HyperSmooth, une mise en œuvre du calcul et de la visualisation de potentiels dans un contexte de cartographie interactive. L'objectif étant de fournir un outil accessible à des utilisateurs via le Web, tout en assurant la sécurité et la confidentialité de ces données. Nous montrons que la méthode est adaptée pour étudier la propagation spatiale de phénomènes sociaux, environnementaux, ou économiques. Nous soulignons que le paramétrage d'une telle analyse peut s'avérer délicat, et nécessite donc de développer un outil adapté afin de rendre les choix plus compréhensibles pour un utilisateur néophyte.

La réalisation d'un tel prototype nécessite une réflexion avancée et coordonnée sur l'architecture, les modalités d'optimisation des calculs, et de paramétrage de l'analyse pour offrir une visualisation interactive. En effet, le coût du calcul, et les contraintes de confidentialité sur les données nous placent face à un défi technique.

Le bilan de notre réalisation s'avère positif quant au choix de l'architecture distribuée : répartition des calculs sur un serveur parallèle d'un coté, visualisation et paramétrage de l'analyse sur un client Java intelligent de l'autre, les deux parties étant connectées via un protocole de plus en plus répandu et présentant une accessibilité et sécurité maximale : SOAP, couplé avec un cryptage SSL. De plus les calculs sont accélérés grâce à l'utilisation d'une méthode de *cut-off* algébrique, et une tabulation des distances orthodromiques.

Une question d'ordre théorique et algorithmique, porte sur l'introduction d'autres distances, et l'extension de la méthode de façon à illustrer les propriétés anisotropiques de l'espace géographique réel. Comme l'a montré une étude des phénomènes de tempêtes, aux flux très orientés (Boulier, 2003), il est intéressant de développer sur la base de la méthode générale une vision anisotropique de l'espace. Pour quantifier le potentiel en un lieu, il s'agit de tenir compte à la fois de la distance de la source d'information au point à estimer, mais aussi de son orientation dans l'espace. Les corrélations entre les potentiels ainsi déterminés et les dégâts constatés sont très nettes. Cette modification de la méthode générale permet d'intégrer de nouvelles formalisations des mouvements dans l'espace (vents dominants, courant marin, etc.) complétant celles qui sont associées à la fonction d'interaction spatiale.